\documentclass[graybox]{svmult}

\usepackage{type1cm}        
\usepackage{makeidx}         
\usepackage{graphicx}        
\usepackage{multicol}        
\usepackage[bottom]{footmisc}

\usepackage{newtxtext}       %
\usepackage{newtxmath}       

\usepackage{amsfonts}
\usepackage{amsmath}

\makeindex             

\begin{document}
\begin{center}
\LARGE{ \textbf{A comparative study of ecological networks using spectral projections of normalized graph Laplacian}} \\

\vspace{0.5cm}

\normalsize Shashankaditya Upadhyay$^{1}$, Sudeepto Bhattachrya$^{2}$ \\

\vspace{0.5 cm}

Department of Mathematics, School of Natural Sciences, Shiv Nadar University, Gautam Buddha Nagar, Uttar Pradesh 201 314, India \\ ${}^{1}$ Email: su348@snu.edu.in, ${}^{2}$ Email: sudeepto.bhattacharya@snu.edu.in

\vspace{1cm}
\end{center}
%
%

\abstract*{Ecological networks originating as a result of three different ecological processes are examined and cross-compared to assess if the underlying ecological processes in these systems produce considerable difference in the structure of the networks. Absence of any significant difference in the structure of the networks may indicate towards the possibility of a universal structural pattern in these ecological networks. The underlying graphs of the networks derived by the ecological processes, namely host-parasite interaction, plant pollination and seed dispersion are all bipartite graphs and thus several algebraic structural measures fail to distinguish between the structure of these networks. In this work we use weighted spectral distribution (WSD) of normalized graph Laplacian, which have been effectively used earlier to discriminate graphs with different topologies, to investigate the possibility of existence of structural dissimilarity in these networks. Graph spectrum is often considered a signature of the graph and WSD of the graph Laplacian is shown to be related to the distribution of some small subgraphs in a graph and hence represent the global structure of a network. We use random projections of WSD to $\mathbb{R}^{2}$ and $\mathbb{R}^{3}$ and establish that the structure of plant pollinator networks is significantly different as compared to host-parasite and seed dispersal networks. The structures of host parasite networks and seed dispersal networks are found to be identical. Furthermore, we use some algebraic structural measures in order to quantify the differences as well as similarities observed in the structure of the three kinds of networks. We thus infer that our work suggests an absence  of a universal structural pattern in these three different kinds of networks.}

\abstract{Ecological networks originating as a result of three different ecological processes are examined and cross-compared to assess if the underlying ecological processes in these systems produce considerable difference in the structure of the networks. Absence of any significant difference in the structure of the networks may indicate towards the possibility of a universal structural pattern in these ecological networks. The underlying graphs of the networks derived by the ecological processes, namely host-parasite interaction, plant pollination and seed dispersion are all bipartite graphs and thus several algebraic structural measures fail to distinguish between the structure of these networks. In this work we use weighted spectral distribution (WSD) of normalized graph Laplacian, which have been effectively used earlier to discriminate graphs with different topologies, to investigate the possibility of existence of structural dissimilarity in these networks. Graph spectrum is often considered a signature of the graph and WSD of the graph Laplacian is shown to be related to the distribution of some small subgraphs in a graph and hence represent the global structure of a network. We use random projections of WSD to $\mathbb{R}^{2}$ and $\mathbb{R}^{3}$ and establish that the structure of plant pollinator networks is significantly different as compared to host-parasite and seed dispersal networks. The structures of host parasite networks and seed dispersal networks are found to be identical. Furthermore, we use some algebraic structural measures in order to quantify the differences as well as similarities observed in the structure of the three kinds of networks. We thus infer that our work suggests an absence  of a universal structural pattern in these three different kinds of networks.}

\section{Introduction}
\label{sec:1}

Network theory has been extensively used in the recent years to study the interactions and relations between individual components of a complex system [1]. Networks comprise generic representation of complex systems in which the underlying topology is a graph such that the various components of the systems are labelled as vertices and the interaction between these components is represented as edges. A formal definition of networks can be found in [2]. Networks thus provide an effective approach to mathematically model empirical data from real world problems where the relationship between given components is of importance. Network theory essentially analyses the structural and functional properties of such real world network models to reveal properties of the underlying complex system that may not be known previously.

Ecological networks can be broadly classified as food web networks [3] and networks of ecological connectivity of certain species [2, 4-6]. In a food web network, vertices represent species or a group of species and edges represent the relation of carbon flow between the species. In a food web network the carbon flow between the species is usually due to predation but this is not always the case. In this work we study food web network representations of three different ecological process such that the carbon flow in each of these three kind of networks is not a predator-pray relationship. These three different networks are enlisted as host-parasite networks, plant-pollinator networks and seed-dispersal networks. 

A host-parasite network represents an ecosystem that comprises of some host species and parasite species that live and reproduce within the host species such that host species form their primary source of nutritional provision [7]. Typically, parasitic species of a host species do not share a host-parasite relationship among themselves [8]. Plant-pollinator network comprises of plant and pollinator species such that a plant species is considered to be connected in the network to a pollinator species if the pollinator species pollinates the given plant species [9,10]. Seed-dispersal networks represent the flow of ecological information in an ecosystem comprising of a set of plant species such that the ecological information (seeds) is dispersed by another set of species i.e. the dispersing species [11,12]. By the virtue of the nature of ecological interactions in these three different kinds of networks, the underlying graphs of these networks are all bipartite graphs.

It is generally assumed that the complexity of such food web networks is captured in some simple algebraic measures such as connectance [13] and in literature the structure of these networks is often presumed to be similar to each other [14,15]. In particular, there has been no study based on spectral graph theory that attempts to distinguish between the structure of these networks. The primary objective of this work is to employ methods developed recently in the field of spectral graph theory to analyse if there is any considerable difference in the structure of the three kind of ecological networks mentioned earlier i.e host-parasite networks, plant-pollinator networks and seed-dispersal networks. In case it is found that there is no significant difference in the structure of these networks then we can assume that there possibly is a universal structural pattern in these networks which may be resulting from the underlying ecological processes.

Spectra of a graph is often considered as a signature of the graph [16,17]. In the current study, we use the applications of weighted spectral distribution (WSD) of the normalized graph Laplacian to discriminate between the structure of the aforementioned three kind of networks [18]. Random projections of weighted spectral distribution have been shown to effectively discriminate between graphs that have different topologies [19]. We further use network motifs, which are thought to be as simple building blocks of large complex networks, to verify if the findings of the spectral methods are consistent and whether these findings can be adequately quantified.

\section{Materials and methods} 
\label{sec:2}

In this study we choose five different networks each from host-parasite ecosystems [20-23] and plant-pollinator ecosystems [24-28] and choose four different networks from seed-dispersal ecosystems [29-32]. All of these networks are available in public domain and can be accessed at \url{https://icon.colorado.edu/}. In addition to their well established role in discriminating graphs with different topologies, a reason for using graph spectra to study these networks emanate from the fact that the underlying graph in these three kinds of networks are all bipartite graphs and thus traditional algebraic measures such as transitivity or clustering coefficient cannot be used to quantify the difference in the topologies of these graphs. 

\subsection{Weighted spectral distribution}

The weighted spectral distribution (WSD) is a spectral measure based on the spectra of normalized graph Laplacian matrix of a graph. Given the adjacency matrix $A$ of a graph $G$, the normalized graph Laplacian $L$ of $G$ can be defined as 
\begin{equation}
L = I - D^{-\frac{1}{2}}AD^{-\frac{1}{2}}\;,
\end{equation}
where $I$ is the identity matrix and $D$ ia a diagonal matrix with entries as the degree of vertices. If $\lambda_{i} i = 0, \dots n-1$ are the eigenvalues of the normalized graph Laplacian then it is known that $0 = \lambda_{0} \leq \lambda_{1}, \dots, \leq \lambda{n-1} \leq 2$ and equality on the upper bound holds iff the graph is bipartite [CN].

If we consider $K$ bins, then a function $\omega(G,N)$ on graph $G$ can be defined as:
\begin{equation}
\omega(G,N) = \sum_{k\in K} (1-k)^{N} f(\lambda = k)\;,
\end{equation}
where $N$ can be chosen as $\{2, 3, \dots\}$ and $f$ is the eigenvalue distribution of the normalized graph Laplacian of $G$.

The elements of $\omega(G,N)$ form the \textit{weighted spectral distribution} that bins the $n$ eigenvalues of the normalized graph Laplacian as:
\begin{equation}
WSD : G \Rightarrow {\mathbb{R}}^{|K|} \{ k \in K : ({(1-k)}^{N} f(\lambda = k))\}\;.
\end{equation}

The structure of a graph is related to WSD as given by the following theorem:

\begin{theorem}
The eigenvalues $\lambda_{i}$ of the normalized Laplacian matrix for an undirected network are related to the closed random walk probabilities as:
\begin{equation}
\sum_{i} {(1-\lambda_{i})}^{N} = \sum_{C} \frac{1}{d_{u_{1}} d_{u_{2}} \dots d_{u_{N}}}\;,
\end{equation}
where $N$ is the length of the random walk cycles, $d_{u_{i}}$ is the degree of vertex $u_{i}$ and $u_{1} \dots u_{N}$ denotes a closed walk from node $u_{1}$ of length $N$ ending at node $u_{N}$ such that $u_{1} = u_{N}$. Here the summation is over all possible closed walks $C$ of length $N$.
\end{theorem}

Thus the left hand side of (4) is related to WSD while the right hand side of (4) is related to distribution of small subgraphs in a graph as given by closed random walks of length $N$. For the purpose of analysis in this work, we choose $N$ as four because the corresponding WSDs in this case are related to closed random walks of length four.The closed random walk of length three are precisely the $3-cycles$ in a simple graph which are absent in bipartite graphs. Thus value of $N$ as three is not chosen for analysis.

\subsection{Bin selection for WSD}

Bins in WSD are assigned such that for a given value of $N$ the sum of weighting in each bin is equal. The weighting in WSD is expressed as:
\begin{equation}
w(x) = (1-x)^{N}\;,
\end{equation}
where $w(x)$ can be thought of as a function that assigns a weight to an eigenvalue of normalized graph Laplacian at $x$. The equality in the sum of weighting in each of the $K$ bins is achieved by solving the integral equation
\begin{equation}
\int_{k_{i}}^{k_{i+1}} w(x)dx = \int_{k_{j}}^{k_{j+1}} w(x)dx\;,
\end {equation}
for all $i, j$. This gives an equal weight of the function $w(x)$ in any pair of given bins $i \in (k_{i}, k_{i+1})$ and $j \in (k_{j}, k_{j+1})$.

\subsection{Random projections of WSD}

Random projection is a general data reduction method which is often used to reduce a high-dimensional data to low-dimensional data for the ease of computations and interpretations. Random projection of WSD has been used effectively in [19] to differentiate between the structure of graphs with different topologies. In order to distinguish $n$ graphs using WSD, consider a matrix $X \in \mathbb{R}^{n\times |K|}$ of WSDs of $n$ graphs with $K$ bins. We obtain a matrix $Y \in \mathbb{R}^{n\times d}$ by multiplying the matrix $X$ with a random projection matrix $R \in \mathbb{R}^{|K|\times d}$, where the elements of $R$ are drawn from a standard normal distribution. Thus we have
\begin{equation}
Y = XR\;,
\end{equation}
such that $R \sim N(0,1)$. The rows of $R$ in expectation form orthogonal vectors as they are normally distributed independent variables with zero correlation. Also the norm of the vectors is $1$ an thus $R$ forms a reduced basis in the original data. 

In the current study a total of fourteen networks from the three different ecosystems have been used to create a data matrix of WSDs. This data matrix is then projected to $\mathbb{R}^{2}$ and $\mathbb{R}^{3}$ using the random projection method described here so that the difference in the structure of these networks can be established using visual inspection of resulting plots.

\subsection{Motifs in networks}

Motifs in a network are subgraphs that are present in the network with a relatively higher frequency as compared to a random network [33]. The relative frequency of different motifs present in the network gives information about the local structure of the network [34-36]. A total of six subgraphs of order four with at least three edges are possible in an undirected graph. These are shown in Fig. 1. In a bipartite graph, cycles of odd length are absent [37]. Thus the only subgraphs that can be found in a bipartite graph are subgraphs (a), (b) and (d) shown in Fig. 1. These subgraphs are commonly known as $claw$, $3-path$ and $4-cycle$ respectively.

\begin{figure}[h]
\sidecaption
\includegraphics[scale=.64]{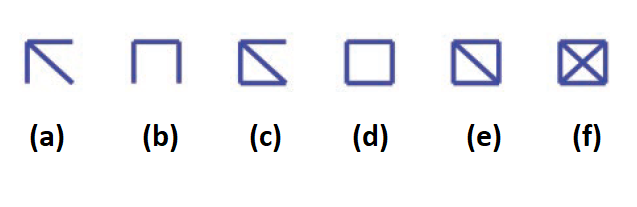}
%
%
\caption{Subgraphs of order four possible in a simple undirected graph. In a simple bipartite graph, subgraph (c), (e) and (f) are absent.}
\label{fig:1}       
\end{figure}

The number $c_{4}$ of $4-cycles$ in a graph is given by:
\begin{equation}
8 c_{4} = tr(A^{4}) + tr(A^{2}) - 2 \sum_{i} d_{i}^{2}\;,
\end{equation}
where, $A$ is the adjacency matrix of the graph, $tr(X)$ is the trace of a matrix $X$ and $d_{i}$ is the degree of a vertex $i$ in the graph $G$.

Also the number of $3-paths$ (subgraph shown in Fig. 1 (b)) in a graph can be calculated using the path matrix $P_{3}$ of the graph $G$. The path matrix is given by:
\begin{equation}
P_{3} = A^{3} - diag(A^{2})A - A diag(A^{2}) + A \times A^{T} - diag(A^{3})\;,
\end{equation}
where $A^{T}$ is the transpose of $A$, diag{A} is the matrix formed by the diagonal elements of $A$ and $\times$ is the element-wise matrix multiplication. An element $(i,j)$ in path matrix $P_{3}$ represents the number of $3-paths$ between vertex $i$ and $j$.

\section{Results}

A summery of the networks used in this study is presented here in Table 1.

\begin{table}
\caption{Summery of order, size and conectance (edge density) of each network.}
\label{tab:1}       
%
%
\begin{tabular}{p{1cm}p{3cm}p{2.2cm}p{2.2cm}p{2.2cm}p{1.8cm}}
\hline\noalign{\smallskip}
S. No. & Network & number of vertices & number of edges & connectance & reference\\
\noalign{\smallskip}\svhline\noalign{\smallskip}
1 & host-parasite 1 (HP1) & 50 & 91 & 0.0222 & [20]\\
2 & host-parasite 2 (HP2) & 175 & 384 & 0.0052 & [21]\\
3 & host-parasite 3 (HP3) & 66 & 114 & 0.0177 & [22]\\
4 & host-parasite 4 (HP4) & 70 & 158 & 0.0127  & [22]\\
5 & host-parasite 5 (HP5) & 130 & 316 & 0.0063 & [23]\\
6 & plant-pollinator 1 (PP1) & 371 & 927 & 0.0022 & [24]\\
7 & plant-pollinator 2 (PP2) & 159 & 204 & 0.0099 & [25]\\
8 & plant-pollinator 3 (PP3) & 174 & 623 & 0.0032 & [26]\\
9 & plant-pollinator 4 (PP4) & 115 & 184 & 0.0109 & [27]\\
10 & plant-pollinator 5 (PP5) & 114 & 167 & 0.0120 & [28]\\
11 & seed-dispersal 1 (SD1) & 40 & 119 & 0.0169 & [29]\\
12 & seed-dispersal 2 (SD2) & 23 & 33 & 0.0625 & [30]\\
13 & seed-dispersal 3 (SD3) & 55 & 211 & 0.0095 & [31]\\
14 & seed-dispersal 4 (SD4) & 26 & 46 & 0.0444 & [32]\\
\noalign{\smallskip}\hline\noalign{\smallskip}
\end{tabular}
\vspace*{-12pt}
\end{table}

Bin selection for WSDs was performed using the method described in section ~\ref{sec:2}. A total of twenty bins were selected for the current study with equal weight of function $w(x) = (1-x)^{4}$ in each bin. Thereafter the weighted spectral distribution for each of the fourteen networks originating from the three different ecosystems was calculated using the bins. The WSDs were plotted subsequently. The plots of WSDs for host-parasite networks, plant-pollinator networks and seed dispersal networks are given as Fig. 2, Fig. 3 and Fig. 4 respectively.

\begin{figure}[h]
\sidecaption
\includegraphics[scale=.32]{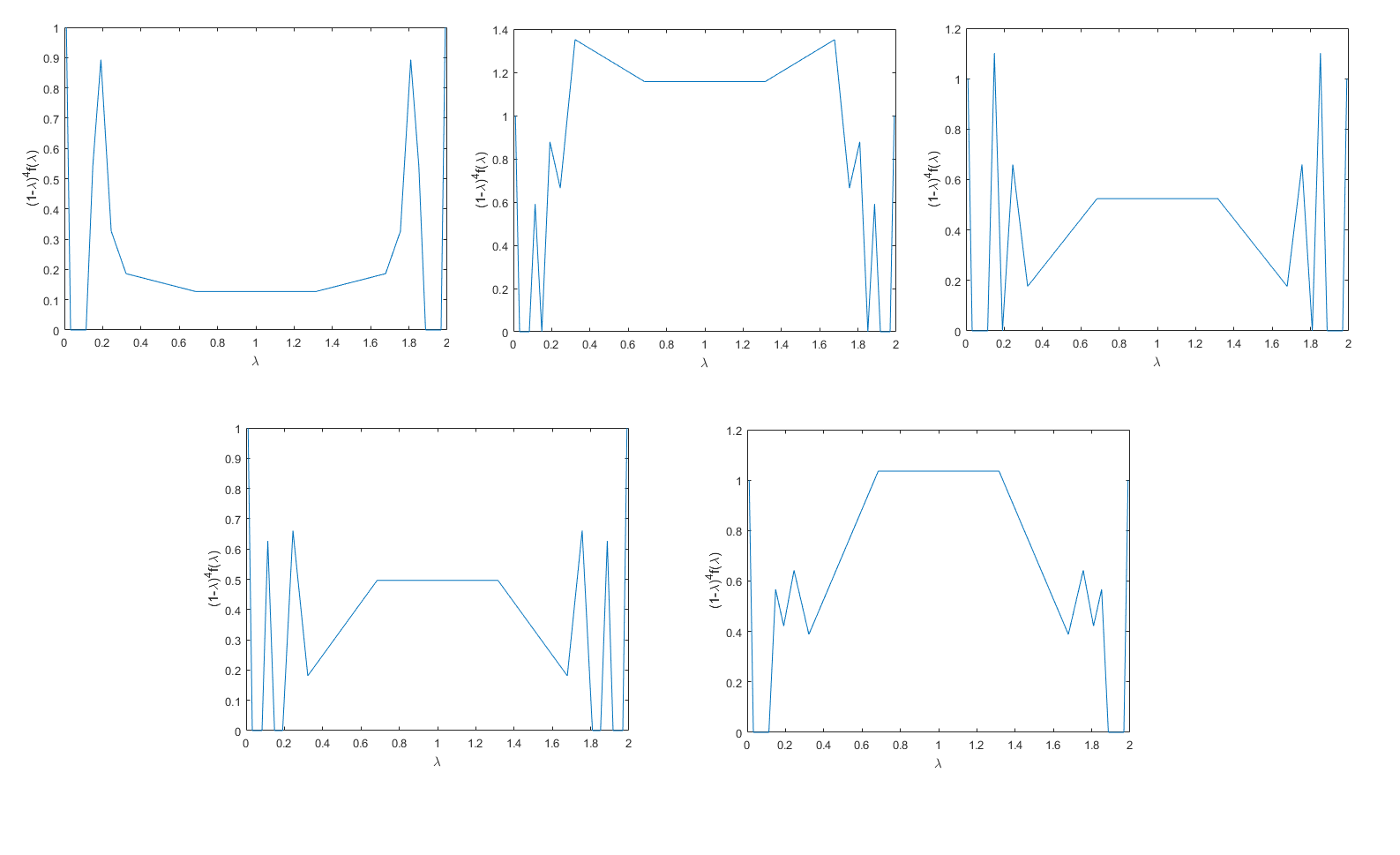}
%
%
\caption{Weighted spectral distributions of host-parasite networks.}
\label{fig:2}       
\end{figure}

\begin{figure}[h]
\sidecaption
\includegraphics[scale=.32]{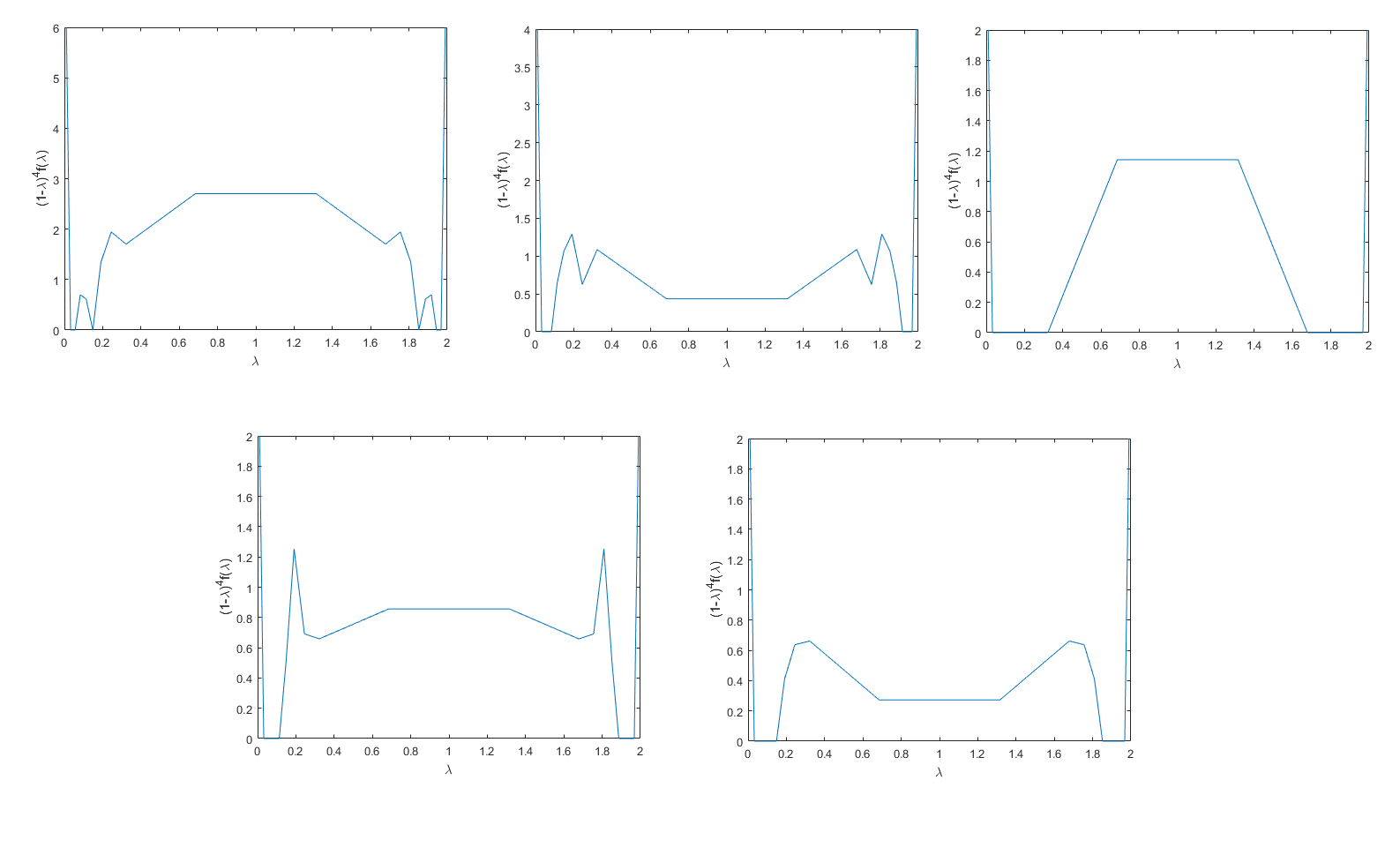}
%
%
\caption{Weighted spectral distributions of plant-pollinator networks.}
\label{fig:3}       
\end{figure}

\begin{figure}[h]
\sidecaption
\includegraphics[scale=.40]{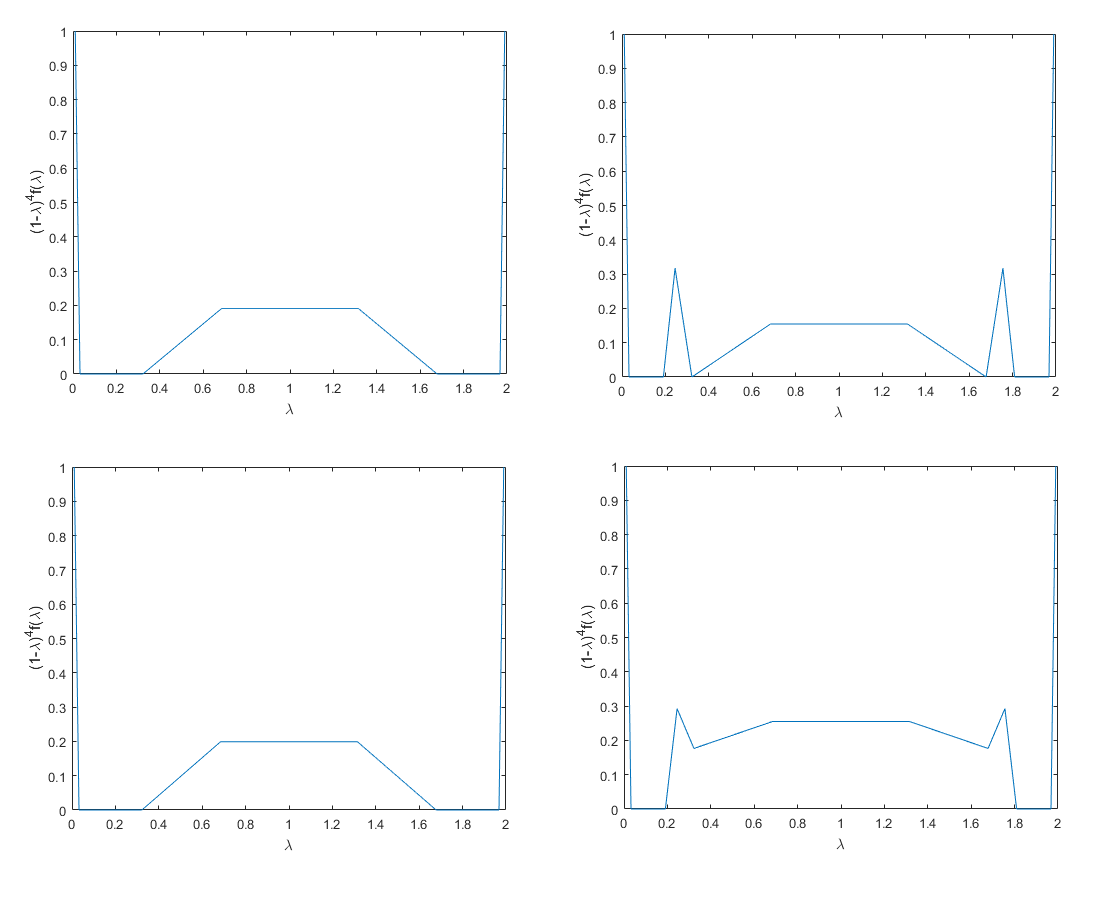}
%
%
\caption{Weighted spectral distributions of seed-dispersal networks.}
\label{fig:4}       
\end{figure}

The WSDs of there networks were projected to $\mathbb{R}^{2}$ and $\mathbb{R}^{3}$ using the random projection method. These projections are shown using plotting the points in $\mathbb{R}^{2}$ and $\mathbb{R}^{3}$ and the resultant plots are given here as Fig. 5 and Fig. 6 respectively.

\begin{figure}[h]
\sidecaption
\includegraphics[scale=.8]{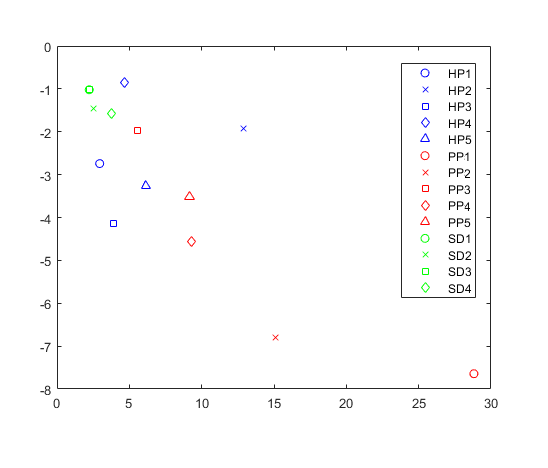}
%
%
\caption{Random projection of weighted spectral distributions to $\mathbb{R}^{2}$. The axis in this graph is irrelevant, only the separation between the points is of significance.}
\label{fig:5}       
\end{figure}

\begin{figure}[h]
\sidecaption
\includegraphics[scale=.29]{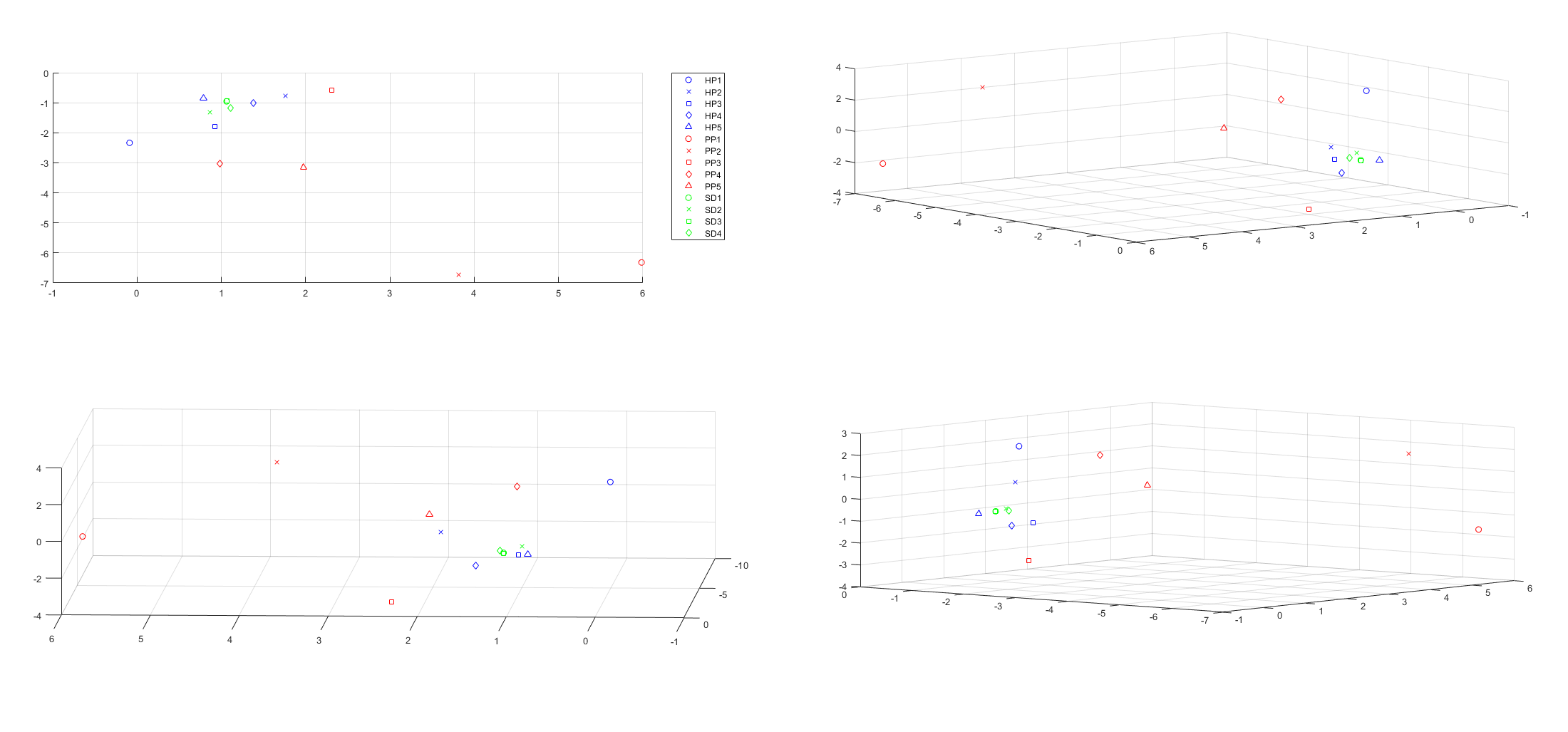}
%
%
\caption{Random projection of weighted spectral distributions to $\mathbb{R}^{3}$. The axis in this graph is irrelevant, only the separation between the points is of significance.}
\label{fig:6}       
\end{figure}

The number of occurence of each of the subgraph $claw$, $3-path$ and $4-cycle$ was calculated in each of the network and the findings are summerized in Table 2.

\begin{table}
\caption{Number of occurence of subgraphs $claw$, $3-path$ and $4-cycle$ in each network.}
\label{tab:2}       
%
%
\begin{tabular}{p{1cm}p{3cm}p{3cm}p{3cm}p{3cm}}
\hline\noalign{\smallskip}
S. No. & Network & number of $claws$ & number of $3-paths$ & number of $4-cycles$  \\
\noalign{\smallskip}\svhline\noalign{\smallskip}
1 & host-parasite 1 (HP1) & 1745 & 2387 & 237\\
2 & host-parasite 2 (HP2) & 20921 & 31666 & 1914\\
3 & host-parasite 3 (HP3) & 2643 & 3467 & 296\\
4 & host-parasite 4 (HP4) & 3879 & 7251 & 704\\
5 & host-parasite 5 (HP5) & 16188 & 25973 & 1923\\
6 & plant-pollinator 1 (PP1) & 194440 & 195470 & 10584\\
7 & plant-pollinator 2 (PP2) & 84943 & 8148 & 143\\
8 & plant-pollinator 3 (PP3) & 219528 & 164495 & 16470\\
9 & plant-pollinator 4 (PP4) & 18385 & 6436 & 287\\
10 & plant-pollinator 5 (PP5) & 48932 & 6214 & 282\\
11 & seed-dispersal 1 (SD1) & 4976 & 7449 & 1107\\
12 & seed-dispersal 2 (SD2) & 411 & 343 & 34\\
13 & seed-dispersal 3 (SD3) & 15222 & 25862 & 4005\\
14 & seed-dispersal 4 (SD4) & 352 & 673 & 55\\
\noalign{\smallskip}\hline\noalign{\smallskip}
\end{tabular}
\vspace*{-12pt}
\end{table}

The relative frequency of occurence of each of the motifs possible in our networks i.e. $claw$, $3-path$ and $4-cycle$ was calculated and the resultant motif profile of the graph is given as Fig. 7.

\begin{figure}[h]
\sidecaption
\includegraphics[scale=.8]{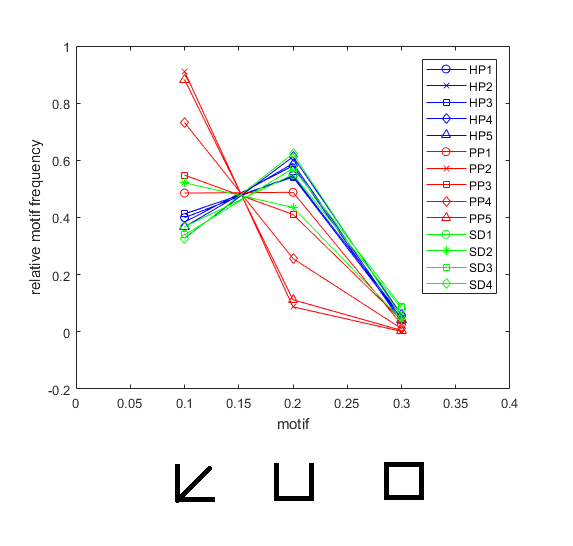}
%
%
\caption{Retaive frequencies of motifs}
\label{fig:7}       
\end{figure}

\section{Discussion and conclusion}

We examined food webs networks originating from three different ecosystems, namely host-parasite, plant-pollinator and seed-dispersal ecosystems. It is often assumed that the structure of these networks can be described by a few simple algebraic measures such as connected. We calculated the value of connectance for each of these networks and observed that there is not any significant difference in the values of connectance across the networks. Thus the observation ab initio indicates towards a possibility of a universal structural pattern in these ecological networks.

However this picture is changed when we consider observations made by visually inspecting the plots that represent the random projection of WSDs of each network to $\mathbb{R}^{2}$ and $\mathbb{R}^{3}$. It is observed that in the plots of random projection of WSDs to $\mathbb{R}^{2}$ and $\mathbb{R}^{3}$, the points that represent host-parasite networks and seed-dispersal networks tend to be clustered together while the points that represent the plant-pollinator networks tend to lie away from this cluster. Since in the plots of random projection of WSDs it is assumed that the points that lie closer to each other are similar in terms of their structure represented by distribution of subgraphs determined by $N$, we can assume that the host-parasite networks are similar to seed-dispersal networks in term of their structure. At the same time, since points that are separated by large distance in a plot of random projection of WSDs are assumed to be graphs that differ in their topology, we may conclude that the structure of plant-pollinator networks is dissimilar as compared to host-parasite and seed-dispersal networks.

To validate our conclusion of a difference in structure between plant-pollinator networks as compared to host-parasite and seed dispersal networks, we enumerated the different subgraphs of order four i.e. the $claw$, $3-path$ and $4-cycle$ possible in the given networks. The reason behind enumerating these subgraphs in the networks is that the observed values are quantifiable and the relative frequency of these subgraphs is known to vary similarly in graphs with similar topologies. We observe that the relative frequency of $claw$ subgraph observed in the plant-pollinator networks is significantly higher as compared to host-parasite and seed dispersal networks. At the same time, the relative frequency of $3-paths$ is relatively lower in plant-pollinator networks as compared to the other two kinds of networks studied here. The values of frequencies of $4-cycles$ are similar in networks but we still observe that these values are collectively lowest in class for the plant-pollinator networks. The relative frequencies of the three subgraphs is found to be similar in host-parasite and seed-dispersal networks. 

Thus we conclude that the plant-pollinator networks are different as compared to host-parasite and seed dispersal networks in terms of their topology. Thus we conclude that a universal structural pattern is absent in the three classes of networks. We also infer that the host-parasite networks are similar to seed-dispersal networks in terms of their structure. 

The absence of a universal structural pattern could be a result of difference in the ecological processes in these ecosystems and the observed difference in the structural pattern open a venue of further research that should be conducted to establish the source of the difference observed in this study. We must conclude however by asserting that use of properties of graph spectra has been been found to effectively differentiate between the structure of these networks and thus provide us with an essential tool to study and cross-compare networks originating from different systems.

\end{document}